\documentclass[12pt]{article}
\usepackage{amssymb}
\usepackage{amsmath}

\newcommand{\be}{\begin{equation}}
	\newcommand{\ee}{\end{equation}}
\def\bea{\begin{eqnarray}}
	\def\eea{\end{eqnarray}}

\newcommand{\bn}{\begin{eqnarray}}
	\newcommand{\en}{\end{eqnarray}}

\newcommand{\p}{\partial}

\newcommand{\nn}{\nonumber}

\newcommand{\no}{\noindent}

\newcommand{\dl}{\delta_{\Lambda}\,}
\newcommand{\dlb}{\delta_{\bar{\Lambda}}}
\newcommand{\vf}{\varphi}
\newcommand{\vft}{\tilde{\varphi}}

\newcommand{\cf}{{\cal F}}

\newcommand{\id}{1\!\!\|}

\def\bea{\begin{eqnarray}}
	\def\eea{\end{eqnarray}}
\def\beq{\begin{eqnarray}}
	\def\eeq{\end{eqnarray}}

\def\bea{\begin{eqnarray}}
	\def\eea{\end{eqnarray}}

\usepackage{xcolor}

\title{\textbf{More on unconstrained descriptions of Higher Spin Massless Particles}}

\begin{document}
	\author{ R. Schimidt Bittencourt \footnote{raphael.schimidt@unesp.br}, D. Dalmazi \footnote{denis.dalmazi@unesp.br}, E.L. Mendonça \footnote{elias.leite@unesp.br}\\
		\textit{{UNESP - Campus de Guaratinguetá - DFI }}} 
	\date{\today}
	\maketitle
	
	\begin{abstract}
	Here we suggest a new local action describing arbitrary integer spin-$s$ massless particles in terms of only two symmetric fields $\vf $ and $\alpha$ of rank-$s$ and $(s-3)$ respectively.  It is an unconstrained version of the Fronsdal theory where the double traceless constraint on the physical field is evaded via a rank-$(s-4)$ Weyl like symmetry. The constrained  higher spin  diffeomorphism is enlarged to full diffeomorphism via the Stueckelberg field $\alpha$ through an appropriate field redefinition. After a partial gauge fixing where the Weyl symmetry is broken while preserving diffeomorphisms, the field equations reproduce, for arbitrary integer spin-$s$, diffeomorphism invariant equations of motion previously obtained via a truncation of the spectrum of the open bosonic string field theory in the tensionless limit. 
	
	In the $s=4$ case we show that the functional integration over $\alpha$
	leads to a unique non local Weyl and diffeomorphism invariant action given only in terms of the physical field $\vf$ whose spectrum is confirmed via an analysis of the analytic structure of the spin-4 propagator for which we introduce a complete basis of projection and transition non local differential operators. We also show that the elimination of $\alpha$ after the Weyl gauge fixing leads to a non local diffeomorphism invariant action previously obtained in the literature.

	\end{abstract}
	\newpage
	
	\tableofcontents
	
	\newpage
	
	\section{Introduction}
	
The subject of higher spin (HS) particles\footnote{The HS particles are understood as particles with spin $s$ such that $s > 2$.}, their existence and mathematical description has a quite long history, see the book \cite{book}, the following review works \cite{snow,rakibur_r,sagnoti_r1,vasiliev_r} and \cite{sagnoti_r}. In the present work, we focus exclusively on free theories of HS fields. Significant challenges are encountered when interactions are introduced.  Although we do have examples of interacting HS models as in string theory (massive particles) and the Vasiliev theory \cite{vasiliev,vasiliev2} (massless), the formulation of interacting HS theories is affected usually by several problems like the propagation of unphysical polarizations \cite{wein}, acausality \cite{velo} and inconsistent coupling to a general gravitational background \cite{aragone}. There are long-standing no-go theorems by Weinberg \cite{wein}, Coleman-Mandula \cite{cm},  Weinberg-Witten \cite{WeinbergandWitten} and its generalization by Porrati \cite{Porrati}, which are still relevant obstacles today. More specifically, long range interactions of massless HS particles are not consistent with HS gauge invariance \cite{wein} and their coupling to the gravitational field contradicts the equivalence principle \cite{Porrati}. In order to avoid the restrictions of the no-go theorems their assumptions must be violated. One can for instance, add masses and have infinitely many HS particles as in string theory, introduce a non vanishing cosmological constant as in the Vasiliev theory to avoid the very restrictive propagation of HS particles in flat space. The no-go theorems and possible exits are nicely reviewed in \cite{bbs}.

 Despite those challenges, progresses have been made, some of them have a string inspired solution. In particular, the lack of causality pointed out by \cite{velo} has found a solution leading to a non-minimal  coupling for spin $s=2$ in \cite{nappi}, see also the solution for $s=3/2$  in \cite{raki32} and for arbitrary spin in \cite{rps}. Another important direction is the construction of cubic vertices of HS particles, even in flat space, which has been put forward via different field theory methods like the Noether approach  \cite{berends,mmr} and the ligh-cone \cite{bbl, metsaev}, it has also benefited from string theory  \cite{sagnoti_r}. In the case of massless HS particles, our main concern here, it is a bit surprisingly that the progresses made via pure field theory methods agree with the string inspired ones since the usual  field theory approach to free massless HS particles (Fronsdal actions \cite{Fronsdal:1978rb})  is a bit different from  the tensionless limit of open string field theory. Our work lies exactly in the connection of the two approaches for free theories.

It turns out that even at free level the HS models are already complicate. A tensor description of a integer spin-$s$ particle requires a rank-$s$ symmetric tensor or a rank-s symmetric spinor tensor for the fermionic case of spin-$(s+1/2)$. As we increase the rank, the more components we have in total and the number of irrelevant entries that must be eliminated increases accordingly. It is natural to use covariant constraints to get rid of those extra undesired fields for massive HS particles. Consistent constraints involving traces and derivatives of symmetric fields have been introduced long ago, see the earlier works \cite{dirac,fp}. Later on, explicit local actions have been successfully formulated for free massive bosons  \cite{shb} and massive fermions \cite{shf} of arbitrary integer and half integer spins. For bosons one makes use of a set of symmetric traceless fields of rank $s,s-2,s-3,\cdots, 0$ or $1$ according to even or odd $s$.  

In the massless limit ($m \to 0$) of \cite{shb} one is left with the rank-$s$ and rank-$(s-2)$ traceless fields which can be combined into only one rank-$s$ field which must be double traceless for $s\ge 4$. Few years after \cite{shb}, Fronsdal \cite{Fronsdal:1978rb}, based on the massless limit of  \cite{shb} , suggested compact and elegant  actions of second order in derivatives for arbitrary integer   massless spin-$s$ particles in terms of only one rank-$s$ tensor   satisfying the double traceless constraint. Accordingly, the massless model is invariant for $s\ge 3$ under a constrained (traceless) HS sort of diffeomorphism. 

Both constraints, on the field itself and on the gauge parameter are not natural from the point of view of a possible HS geometry \cite{wit}. Moreover, they do not appear in the tensionless limit of the open bosonic string field theory where one obtains, for the first Regge trajectory, a reducible system of massless particles of spins $s, s-2,s-4, \cdots 0$ or $1$ described by unconstrained symmetric fields. The corresponding action is invariant under unconstrained (traceful) Diff's, see \cite{os,bengt}. One might speculate that the absence of constraints is an important ingredient behind the selfconsitency of the HS interactions in strings. It is therefore interesting to try to lift those constraints in the irreducible description of spin-s particles and connect the Fronsdal theory \cite{Fronsdal:1978rb} with the reducible unconstrained HS theory coming from string field theory. 

A step toward the reduction of the string spectrum to only spin-s modes has been taken by the authors of \cite{pt,bpt} who have used the BRST technique as in \cite{os,bengt}. Later, the BRST approach was very much improved, see \cite{buch_npb2007}, and the number of symmetric tensor fields for an unconstrained off-shell formulation of massless spin-$s$ bosons, of second order in derivatives, has been reduced down to six fields: four fields
of ranks $s, (s-1), (s-2), (s-3)$ and two Lagrange multipliers of rank-$(s-2)$ and $(s-4)$. Here on the other hand,  we are closer to the approach  started in  \cite{fs1}, \cite{fscqg} and \cite{st} where a ``triplet'' of equations coming from string theory in the tensionless limit, describing massless particles of spin $s, s-2, s-4, \cdots, 0$ or $1$, has been truncated
to an extension of the Fronsdal equation containing only spin-s modes. The resulting equations of motion are invariant under unconstrained Diff. The truncation is implemented via a clever constraint which  introduces in the Fronsdal theory only one extra field of the Stueckelberg type and rank-$(s-3)$, called $\alpha$ field. The reduction is carried out  at the level of the equations of motion only. Later, the addition of a rank-$(s-4)$ Lagrange multiplier, the $\beta$ field, has led the authors of \cite{fs2} to a ``minimal''  local action \cite{fms} describing spin-$s$ massless particles in fully unconstrained form in terms of only three fields $(\vf,\alpha,\beta)$ of rank $s, (s-3)$ and $(s-4)$ respectively. The work of \cite{fs2, fms} also includes the case of fermions and AdS backgrounds.  

Here we show that the number of fields can be further reduced to only two fields $(\vf,\alpha)$ by trading the $\beta$ field into a rank-$(s-4)$ Weyl like  gauge symmetry and introducing $\alpha$ by a field redefinition. In section 2 we revisit the Diff invariant model of \cite{fs2} from our point of view in order to pave the way for our arbitrary integer spin-s model suggested in section 3. In section 4 we work out explicitly the $s=4$ case. In particular, we functionally integrate over the compensator $\alpha$ and deduce a nonlocal action invariant under a rank-$(s-4)$ Weyl symmetry plus unconstrained Diff.  The action correctly describes only spin-4 massless polarizations. The  particle content is checked via an analysis of the analytic structure of the spin-4 propagator making use of a basis of nonlocal differential operators acting on symmetric rank-4 fields displayed in Appendix A. After a partial gauge fixing of the local Lagrangian where the Weyl symmetry is broken leaving only the unconstrained Diff symmetry, we show that the elimination of the $\alpha$ field leads to the same unique spin-4 Diff invariant nonlocal action obtained in \cite{fms}. For completeness we analyse the particle content of a  more general two parameters non local Diff invariant theory containing other classes of Weyl symmetries.

 In section 5 we draw our conclusions and in appendix B we prove a simple theorem regarding the Weyl invariance of our model.

	\section{Revisiting the minimal unconstrained local model}\label{sec:vb-fronsdal}

 \hspace{0.5cm} Our starting point is the Fronsdal theory \cite{Fronsdal:1978rb} for massless bosonic spin-$s$ particles in flat Minkowski  spacetime of $D$ dimensions\footnote{We work in flat Minkowski spacetime with $\eta = \text{diag}(-,+,+,\dots,+)$. Symmetrizations are minimal (without weights), e.g., $T_{(\mu\nu)} = T_{\mu\nu} + T_{\nu\mu}$, except in Appendix-A. We use the compact notation from \cite{fs1}, where symmetrizations are implicit: $\partial^k \vf = \partial_{(\mu_1}\cdots \partial_{\mu_k} \vf_{\mu_{k+1} \dots \mu_{k+s})}$ and $\eta^k \vf = \eta_{(\alpha_1\beta_1}\cdots \eta_{\alpha_k\beta_k} \vf_{\mu_{1} \dots \mu_{s})}$. The dotted terms $\partial \cdot$ and $\eta \cdot$ represent contractions: $(\partial \cdot)^k \vf = \partial^{\mu_1} \cdots \partial^{\mu_k} \vf_{\mu_1 \dots \mu_s}$ and $(\eta \cdot)^k \vf = \eta^{\mu_1 \mu_2} \cdots \eta^{\mu_{2k}} \vf_{\mu_1 \dots \mu_s} \equiv \vf^{[k]}$. Primes denote lower $k$ traces, e.g., $\vf^{[2]} = {\vf}''$. A single bar indicates a traceless field ($\eta^{\mu_1\mu_2} \bar{\vf}{\mu_1 \dots \mu_s} = 0$), and two bars indicate a doubly traceless field ($\eta^{\mu_1\mu_2} \eta^{\mu_3 \mu_4} \bar{\bar{\vf}}{\mu_1 \dots \mu_s} = 0$).
 	
 }. In terms of a totally symmetric rank-$s$ gauge field $\vf_{\mu_1 ... \mu_s}= \vf_{(\mu_1 ... \mu_s)}$ we have 
	
	\be S_F[\vf]= \frac{1}{2} \int \, d^Dx\,\, \vf \, {\cal G}(\vf) = \frac{1}{2} \int \, d^Dx\,\, \vf \, \left\lbrack \cf - \frac 12 \eta\, {\cal F'} \right\rbrack  \label{sf}\ee
	
	\no where the higher spin  Einstein-like tensor, the Ricci-like higher spin curvature (Fronsdal tensor) and its trace are given respectively by:

	\bea {\cal G}_{\mu_1 ... \mu_s}(\vf)&\equiv &{\cal F}_{\mu_1 ... \mu_s}(\vf)-\frac{1}{2}\eta_{(\mu_1\mu_2}{\cal F'}_{\mu_3 ... \mu_s)}(\vf).\label{einstein} \\
 {\cal F}_{\mu_1 ... \mu_s}(\vf) &=& \Box \vf_{\mu_1 ... \mu_s}-\p_{(\mu_1}\p^{\alpha}\vf_{\mu_2 ... \mu_s) \alpha}+\p_{(\mu_1}\p_{\mu_2}\vf'_{\mu_3 ... \mu_s)} \label{fronsdal} \\ 
{\cal F'} &=& 2 \, \Box \vf' - 2 \, \partial \cdot\partial\cdot \vf + \partial(\partial \cdot \vf') + \partial^2 \vf'' \label{fprime}  .\eea

Neglecting total derivatives, the functional differential of (\ref{sf}) is given by

\be \delta\, S_F=  \int \, d^Dx\, \left\lbrack \delta\, \vf \, \left( \cf - \frac 12 \eta\, {\cal F'} \right)  + \frac 14 \delta\, \vf \, \eta\, \partial ^2\vf'' - \frac 32 \binom{s}{4}\, \delta\, \vf'' \, \partial\cdot \partial\cdot \vf' \, \right\rbrack \quad .\label{deltasf}\ee

\no The last two terms show up because of the non-Hermiticiy of the second order differential operator in $S_F$ which, on its turn, is a consequence of the double trace term in ${\cal F'}$.  

In the original Fronsdal theory the gauge field is identically (off-shell) doubly traceless $\vf''=0$, consequently the equations of motion of $S_F$ become the Einstein-like equations $ {\cal G} = \cf -  \eta\, {\cal F'}/2 =0$ which after taking traces lead to $\cf =0 $. Those equations altogether with the constraints $\vf''=0$ correctly describe the free dynamics of a spin-s massless particle. As in the spin-1 (Maxwell) and spin-2 (Einstei-Hilbert) cases a key ingredient in order to achieve the correct number of degrees of freedom (2 in $D=4$) is a rank-$(s-1)$ gauge symmetry. Indeed, once we assume $\vf''=0$ it is not difficult to show that $\delta_{\bar{\Lambda}}S_F =0$ where 

	\be \dlb \vf_{\mu_1 ... \mu_s}= \p_{(\mu_1}\bar{\Lambda}_{\mu_2 ... \mu_s)}\quad; \quad \eta^{\mu^{i}\mu^{j}}\bar{\Lambda}_{\mu_1...\mu_{s-1}}=0 \quad;\quad 1\leqslant \lbrace i,j\rbrace \leqslant s-1 \label{cdiff} \quad .\ee 
	
	\no Since the gauge parameter is subjected to the traceless condition we are going to refer to this gauge transformation as {\it  constrained diffeomorphisms} (CDiff). Notice that if we had added a source term ${\cal L}_J = \vf \, J$ to (1), due to $\vf''=0$, only the doubly traceless piece of the rank-s current would contribute, so $J=\bar{\bar{J}}$ without loss of generality. Moreover, due to (\ref{cdiff}) the current does not need to be conserved but rather satisfies $\partial \cdot \bar{\bar{J}} = \eta\, \bar{j} $ where $\bar{j}$ is an arbitrary traceless rank-(s-3) symmetric tensor, which includes the trivial case $\bar{j}=0$.
	
	So in the Fronsdal theory both the field and the gauge parameter are constrained which is quite unnatural as we have mentioned before. In order to see how those constraints are simultaneously lifted by the unconstrained minimal model of \cite{fs2} it is convenient to perform an unconstrained Diff transformation $\delta_{\Lambda}\vf=\partial\, \Lambda$ in the unconstrained version of  $S_F$, namely,
	
	\bea \delta_{\Lambda} S_F &=& \int d^D x \, \Big\{ -3 \binom{s}{4} \vf'' \, \partial \cdot \partial \cdot \partial \cdot \Lambda + \frac{15}2 \binom{s}{5} \, \Lambda'' \partial \cdot \partial \cdot \partial \cdot \, \vf' \label{deltasfdiff} \\ &+& \frac 34 \binom{s}{3} \, \Lambda' \left\lbrack 6\Box\partial\cdot\vf' -4(\partial\cdot\partial\cdot\partial\cdot\vf) + 4\partial(\partial\cdot\partial\cdot \vf') + \Box \partial \vf'' + \partial^2\, \partial\cdot \vf''\right\rbrack \Big\} \, ,\nn\eea
	
	\no where we have used  the (off-shell) anomalous Bianchi identity \footnote{Notice that the first line and the last three terms of the second line of (\ref{deltasfdiff}) disappear for $s=3$ as well as the right hand side of (\ref{bianchi}).}: 
	
	\be \partial \cdot \, \cf - \frac 12 \partial \, \cf' = - \frac 32\, \partial^3 \, \vf'' \quad  . \label{bianchi} \ee
	
	We see that a rank-$(s-1)$ gauge symmetry for the pure Fronsdal action without any extra field requires both constraints $(\Lambda',\vf'')=(0,0)$. We first concentrate on the removal of the constraint on the trace of the gauge parameter. This will be achieved by the introduction of a symmetric rank-$(s-3)$ field $\alpha$. It is a kind of compensator (Stueckelberg) field which must transform according to the trace of the gauge parameter  $\delta_{\Lambda}\alpha = \Lambda' $. In the next subsection we look at the case $s=3$ as a simpler example where there is no need of the double traceless constraint on the gauge field but the basic idea is the same one for $s>3$.

	  \subsection{From CDiff to Diff via a Stueckelberg field ($\alpha$-field)}

	  We can always decompose a symmetric rank-$k$ tensor in its traceless piece plus another piece which only depends on its rank-$(k-2)$ trace. In the spin-3 case the gauge parameter is a rank-2 tensor and its traceless piece can be written as $\bar{\Lambda}_{\mu\nu} = \Lambda_{\mu\nu} -\Lambda'\eta_{\mu\nu}/D $. Therefore, we can write
	 
	\be  \p_{(\mu} \bar{\Lambda}_{\nu\rho)} = \p_{(\mu} \Lambda_{\nu\rho)} - \frac 1D \p_{(\mu}\Lambda'\, \eta_{\nu\rho)} \ee
	
\no which suggests 	
	
	 \be \dl\, \vft_{\mu\nu\rho} =   \dl \left\lbrack \vf_{\mu\nu\rho} - \frac 1D \p_{(\mu}\alpha\, \eta_{\nu\rho)}\right\rbrack \ee
	\no where
	
	\be \dl \vf_{\mu\nu\rho} =  \p_{(\mu} \Lambda_{\nu\rho)} \quad ; \quad \dl \, \alpha = \Lambda'  \quad \to \quad \dl \vft_{\mu\nu\rho} 
	= \p_{(\mu} \bar{\Lambda}_{\nu\rho)} \, . \label{diff3} \ee
	
	The previous formulae show that the CDiff symmetry of the spin-3 Fronsdal action can be enlarged to Diff at the expense of introducing the Stueckelberg  field $\alpha$ via the simple field substitution:

\be \vft_{\mu\nu\rho} =    \vf_{\mu\nu\rho} - \frac 1D \eta_{(\nu\rho}\p_{\mu)}\alpha\, \label{fredefs3}\ee

\no Indeed, the spin-3 Fronsdal action becomes 

\bea  S_F[\vft] &=& \frac{1}{2} \int \, d^Dx\,\, \vft \, {\cal G}(\vft)\nn\\ &=& \int d^Dx\,\,\Big[ \frac{1}{2} \vf \, \Box \, \vf + \frac 32 (\p\cdot \vf)^2 + 3 \vf' \, \p\cdot \p \cdot \, \vf - \frac 32 \vf' \, \Box \, \vf' + \frac 34 (\p \cdot \vf')^2 \nn\\  &+& 3\, \alpha \, \p\cdot \p \cdot \p \cdot \, \vf - \frac 92 \alpha\, \Box \, \p \cdot \, \vf' + \frac{9}{4} \, \alpha \, \Box^2 \, \alpha  \Big] 
.\label{sf3}
\eea

\no The action (\ref{sf3}), to the best we know, has first appeared in \cite{fs1}, though its equations of motion were obtained long before by Schwinger \cite{schwinger}. The action is invariant under unconstrained Diff (\ref{diff3}). The equations of motion $K^{\mu\nu\rho} \equiv \delta  S_F[\vft]/\delta \vf_{\mu\nu\rho}=0 $ become simply ${\cal G}(\vft)=0$ which, after taking traces, lead to

\be {\cal F}(\vft)=0 \quad\quad \rightarrow \quad\quad {\cal F}(\vf)= 3\, \p\,\p\, \p\, \alpha \quad . \label{falpha3} \ee 

 Throughout this work we do not worry about the equations of motion for the Stueckelberg fields, since they follow from the equations of motion for the gauge field due to the unconstrained gauge symmetry. In particular, from $\dl S(\vft) =0$ we derive, see (\ref{diff3}), after an integration by parts, that 
$K=(3/D)\, \p_{\mu}K^{\mu} $ where $ K  \equiv \delta  S_F[\vft]/\delta\alpha $ and $K^{\mu} = \eta_{\nu\rho}K^{\mu\nu\rho}$. Moreover, since the gauge transformations of  $\alpha $ contain no derivatives, the field is said to be  pure gauge, then the theorem proven in \cite{moto1}  applies, even for higher derivative theories, which means that  we can fix the gauge $\alpha =0$ at action level without losing physical content which assures in particular, that we do not need to worry about the  Ostrogradsky ghost \cite{ostro,wood}. In this gauge $ S_F[\vft]= S_F[\vf]$ and the  action becomes the usual spin-3 Fronsdal theory, so the pure spin-3 content is safe\footnote{One can also functionally integrate over $\alpha$ in (\ref{sf3}) and check that the particle content of the corresponding non local action contains only a physical spin-3 massless pole. }. 
 
 Now we move to arbitrary integer spins.  Although, we still have to deal with the lifting of the double traceless constraint for $s\ge 4$, the symmetry enlargement CDiff $\to$ Diff works similarly for any integer spin. The field redefinition (\ref{fredefs3}) can be generalized for arbitrary spin-$s$ once we know the decomposition of a symmetric tensor of  arbitrary rank  into its traceless and remaining trace dependent terms and this is given explicitly in \cite{fms}. Namely, 
 
 \bea \Lambda_{\mu_1\cdots \mu_{s-1}} &=& \bar{\Lambda}_{\mu_1\cdots \mu_{s-1}} + \frac{\eta_{(\mu_1\mu_2}\Lambda'_{\mu_3\cdots \mu_{s-1})}}{D+ 2(s-3)} - \frac{\eta_{(\mu_1\mu_2}\eta_{\mu_3\mu_4}\Lambda''_{\mu_5\cdots \mu_{s-1})}}{[D+ 2(s-3)][D+ 2(s-4)]} + \cdots \nn\\ 
 \Lambda_{\mu_1\cdots \mu_{s-1}} &=& \bar{\Lambda}_{\mu_1\cdots \mu_{s-1}} + f_{\mu_1\cdots \mu_{s-1}} (\Lambda') \quad . \label{lambdas} \eea

\no where, in compact notation, the rank-$(s-1)$ symmetric field $f$ is given by:

\be f (\Lambda') = - \sum_{k=1}^{\left\lbrack \frac{s-1}{2} \right\rbrack} \frac{(-\, \eta)^k \Lambda^{[k]}}{\prod_{j=0}^{k-1}[D+ 2(s-3-j)]}
\quad . \label{fs-1} \ee

\no In agreement with ({\ref{lambdas}), $f$ satisfies:

\be f'(\Lambda') = \Lambda' \quad \rightarrow \quad f''(\Lambda') = \Lambda'' \quad . \label{ftraces} \ee 

\no Introducing a rank-$(s-3)$ symmetric tensor $\alpha$  we have from ({\ref{lambdas}), in compact notation,

\be \p \, \Lambda = \p \, \bar{\Lambda} + \p \, f (\Lambda') \quad \rightarrow \quad \dl \vft = \dl \left\lbrack \vf - \p\, f (\alpha) \right\rbrack \quad , \label{del_lambdas}\ee
\no where
\be  \dl  \vf =  \p\, \Lambda \quad ; \quad \dl  \alpha = \Lambda' \quad \rightarrow \quad \dl  \vft = \p \,\bar{\Lambda} \quad , \label{diffs} \ee

\no which leads to the spin-$s$ generalization of the spin-3 field redefinition (\ref{fredefs3}), i.e.,  

\be \vft = \vf - \p \, f(\alpha) \quad . \label{fredefs} \ee

\no On the other hand, from the very definition of the Fronsdal tensor (\ref{fronsdal}) we can show the known relation ${\cal F}(\vf + \p\, \Lambda) = 
{\cal F}(\vf) + 3\, \p\, \p\, \p\, \Lambda'  $ which altogether with the first of the relations (\ref{ftraces}) with $\Lambda' \to \alpha$ allows us to deduce a key Diff invariant identity for the remaining of this work which holds true for arbitrary integer spin-$s$:

\be {\cal F}(\vft) = {\cal F}\left[\vf - \p\, f (\alpha )\right] = 
{\cal F}(\vf) - 3\, \p\, \p\, \p\, \alpha  \quad  \label{id1} \ee 

\no Moreover, from the double trace of the field redefinition (\ref{fredefs}) and the identities (\ref{ftraces}) with $\Lambda' \to \alpha$ we obtain another important identity also invariant under unconstrained diffeomorphisms,

\be \vft'' = \vf'' -4\, \p \cdot \, \alpha - \p\, \alpha'  \quad . \label{id2} \ee 

\subsection{Lifting the double traceless constraint ($\beta$-field)}

The minimal model of \cite{fs2} lifts the double traceless constraint for $s\ge 4$ via a rank-$(s-4)$ symmetric tensor ($\beta$) which works like a Lagrange multiplier enforcing  the double traceless constraint on shell. It is instructive to look at the following modified Fronsdal model as an intermediate step,

\be S_{\beta} =  \int \, d^Dx\,\, \left\lbrace \frac{\vf}2 \, \left[ \cf (\vf) - \frac 12 \eta\, {\cal F'}(\vf) \right] + 3\, \binom{s}{4} \, \beta \, \vf'' \right\rbrace  \label{sbeta}\ee

\no The $\beta$ and $\vf$ equations of motion lead respectively to 

\bea \vf'' &=& 0 \quad , \label{eq1} \\
{\cal G}(\vf) + \frac{\eta\, \p\,\p\, \vf''}4 + \eta\, \eta \left(\beta - \frac{\p\cdot \p\cdot \vf'}2 \right) &=& 0  \label{eq2} \eea

\no Since ${\cal G''}(\vf)=0$ due to (\ref{eq1}), after taking traces of (\ref{eq2}) we conclude that (\ref{eq1}) and (\ref{eq2}) are equivalent to

\be \beta = \frac{\p\cdot \p\cdot \vf'}2 \quad ; \quad \vf'' =0 \quad ; \quad {\cal F}(\vf) = 0  \quad . \label{eq3} \ee 

\no So we recover the usual Fronsdal equations of motion  and verify that the  $\beta$ field is not an independent degree of freedom. We conclude that  $S_{\beta}$ is equivalent to the Fronsdal theory. In particular, both models are off-shell invariant under CDiff  ($\delta_{\bar{\Lambda}}\vf = \p \, \bar{\Lambda}$), see (\ref{deltasfdiff}) if, in agreement with (\ref{eq3}), we define\footnote{Notice that the specific numerical factor in the $\beta$ term in $S_{\beta}$ has been chosen in order to simplify  (\ref{deltabeta}).}

\be \delta\beta = \p \cdot  \p \cdot  \p \cdot  \, \bar{\Lambda} \quad . \label{deltabeta} \ee. 

\subsection{ The unconstrained  $\alpha\beta$ model}

Now we are ready to define a model describing spin-$s$ particles where both constraints on the gauge parameter and on the gauge field are lifted. Namely,

\be S_{\alpha\beta}[\vf,\alpha , \beta] =  \int \, d^Dx\,\, \left\lbrace \frac{\vft}2 \, \left[ \cf (\vft) - \frac 12 \eta\, {\cal F'}(\vft) \right] + 3\, \binom{s}{4} \, \beta \, \vft'' \right\rbrace  \label{sab}\ee

\no where $\vft$ is defined in ({\ref{fredefs}) and ({\ref{fs-1}). 

Under {\bf unconstrained} Diff's, see  (\ref{deltasf}) with $(\vf,\Lambda) \to (\vft,\bar{\Lambda})$ and (\ref{diffs}), we have 

\be \dl S_{\alpha\beta} = \int \, d^Dx\,\, \left\lbrack 3\, \binom{s}{4} 
\left( \dl \beta - \p \cdot  \p \cdot  \p \cdot  \, \bar{\Lambda}  \right) \, \vft'' \right\rbrack \quad . \label{dsab} \ee

\no Defining $\dl \beta = \p \cdot  \p \cdot  \p \cdot  \, \bar{\Lambda} $ we have $\dl S_{\alpha\beta}=0$. Since the $\alpha$ field is pure gauge, according to the theorem of \cite{moto1} we can fix the gauge $\alpha =0$ at action level. This turns $S_{\alpha\beta}$ into $S_{\beta}$ which demonstrates that $S_{\alpha\beta}[\vf,\alpha , \beta] $ correctly describes spin-$s$ particles with unconstrained Diff and no double traceless requirement on the gauge field. The action $S_{\alpha\beta}$ is almost the same minimal action suggested in \cite{fs2}. For a closer comparison we notice that

\bea S_{\alpha\beta} &=&  \int \, d^Dx\,\, \left\lbrace \frac{\left[\vf - \p\, f (\alpha)\right]}2 \, \left\lbrack {\cal F}(\vft) - \frac 12 \eta\, {\cal F'}(\vft) \right\rbrack + 3\, \binom{s}{4} \, \beta \, \vft''\right\rbrace  \nn\\ 
&=&\int \, d^Dx\,\,  \left\lbrace \frac{\vf }2 \, \left\lbrack {\cal F}(\vft) - \frac 12 \eta\, {\cal F'}(\vft) \right\rbrack + 3\, \binom{s}{4} \, \beta \, \vft'' +  \Delta \, {\cal L} \right\rbrace  
\eea

\no where, neglecting a total derivative, 

\bea \Delta \, {\cal L} &=& -\frac {1}{2[D+2(s-3)]} \left\lbrack \eta^{(\mu_1\mu_2}\p^{\mu_3}\, \alpha^{\mu_4\mu_5\cdots \mu_s)} - \frac{\eta^{(\mu_1\mu_2}\eta^{\mu_3\mu_4}\p^{\mu_5}(\alpha')^{\mu_6\cdots \mu_s)}}{2[D+2(s-4)]} + \cdots \right\rbrack\nn\\
&\times&  \left\lbrack {\cal F}_{\mu_1\cdots \mu_s}(\vft) - \frac{\eta_{(\mu_1\mu_2}{\cal F}'_{\mu_3\cdots \mu_s)}(\vft)}{2} \right\rbrack \nn\\
&=& - \frac 34 \binom{s}{3} \alpha \, [\p \cdot {\cal F}'(\vft)] + \cdots \label{deltal} \eea

\no The dots in the first equality of (\ref{deltal}) stand for the remaining terms of the series (\ref{fs-1}) starting at $\eta^3$ while the dots in the second equality of (\ref{deltal}) contain linear terms on ${\cal F}^{[k]}$   with $k\ge 2$. Since those terms depend upon $\vft^{[k]}$ with $k\ge 2$ it is always possible to get rid of them by taking advantage of the term $\beta\, \vft''$  and redefining $\beta \to \tilde{\beta}$ with $(\tilde{\beta}-\beta)$ being linear on $\alpha$ and its traces such that we can suppress the dots in the second equality of (\ref{deltal}). Using (\ref{id2}) we end up with the same unconstrained\footnote{Although the $\beta$ field enforces $\vft'' =0$ on shell in (\ref{sab}) and  (\ref{sfs}), both actions are called unconstrained since there is no off-shell constraints on the fields nor on the gauge parameter. Moreover, one can in principle  integrate over the pair $(\alpha,\beta)$ and obtain a non local field theory in terms only of the unconstrained rank-s field $\vf$, see \cite{francia_int} for an explicite integration in the $s=4$ case.}  Diff invariant action of \cite{fs2}, namely, 

\be S_{FS} = \int \, d^Dx  \left\lbrace \frac{\vf }2 \, \left\lbrack {\cal A} - \frac 12 \eta\, {\cal A'} \right\rbrack - \frac 34 \binom{s}{3} \alpha \, [\p \cdot {\cal A}'] + 3\, \binom{s}{4} \, \tilde{\beta}\, (\vf'' -4\, \p \cdot \, \alpha - \p\, \alpha') \right\rbrace  \label{sfs}
\ee

\no where, in the notation of \cite{fs2},  $  {\cal A} = \cf (\vft) =  {\cal F}\left[\vf - \p\, f (\alpha )\right] = {\cal F}(\vf) - 3\, \p\, \p\, \p\, \alpha  $ is invariant under unconstrained Diff's. The redefinition $\beta \to \bar{\beta}$ changes the Diff transformation (\ref{deltabeta}) into $\dl \bar{\beta } = \p \cdot  \p \cdot  \p \cdot  \, \Lambda $ which altogether with (\ref{diffs}) leads to $\dl S_{FS} =0$, see \cite{fs2}.

 \section{The unconstrained $\alpha$ model}
 
In this section we build up a local model  $S_{\alpha}[\vf,\alpha] $ describing  massless spin-$s$ particles in terms of only two unconstrained symmetric fields, one ($\vf$) of rank-$s$   and another one ($\alpha$) of rank-$(s-3)$. 
We begin by taking a detour regarding section 2.2 and building up an intermediate action $S_I (\vf )$ where the double traceless condition is assured by a rank-$(s-4)$ Weyl symmetry instead of a Lagrange multiplier $\beta$.  Namely, we define
 
 \be S_I[\vf] = S_F\left\lbrack \phi(\vf)\right\rbrack= \int \, d^Dx\,\, \left\lbrace \frac{\phi}2 \, \left[ \cf (\phi) - \frac 12 \eta\, {\cal F'}(\phi) \right]  \right\rbrace  \label{si}\ee
 
 \no where $\phi(\vf)$, in compact notation with implicit symmetrization,  is given by the finite series
 
 \be \phi (\vf) \equiv \vf + \sum_{n=2}^{\left\lbrack \frac s2 \right\rbrack }(n-1)c_{n-1} \, \eta^n \, \vf^{[n]} \quad , \label{phi1} \ee
 
 \no with 
 
 \bea c_1 &=& - \frac 1{[D + 2(s-3)][D + 2(s-4)]} \label{c1} \\ c_n &=& - \frac{c_{n-1}}{[D + 2(s-3-n)]} \quad ;  \quad n\ge 2 \quad . \label{cn} \eea
 
 \no Although $\vf $ is unconstrained, the coefficients $c_n$ guarantee that $\phi$ is identically  double traceless ($\phi'' =0$), without being  traceless ($\phi' \ne 0$) which makes $\phi$ and consequently $S_I$ automatically invariant under a rank-$(s-4)$ Weyl transformation, $\delta_{\lambda} \phi (\vf) = 0$, as shown in Appendix B, where,
  
 \be \delta_{\lambda}\vf_{\mu_1\cdots s} = \eta_{(\mu_1\mu_2}\, \eta_{\mu_3\mu_4}\,\lambda_{\mu_5\mu_6\cdots \mu_s)} \quad . \label{weylt} \ee

 Since $S_I$ depends upon $\vf$ exclusively via $\phi(\vf)$, we have
 
 \bea \frac{\delta S_I}{\delta\vf_{\mu_1 \cdots \mu_s}} &=&  \frac{\delta\phi_{\nu_1\cdots \nu_s}}{\delta\vf_{\mu_1 \cdots \mu_s}} \, \frac{\delta S_I}{\delta\phi_{\nu_1 \cdots \nu_s}}  
 =  \left\lbrack \delta_{\nu_1}^{\mu_1}\cdots \delta_{\nu_s}^{\mu_s}\,  + \, c_1 \,\frac{ \eta_{(\nu_1\nu_2}\eta_{\nu_3\nu_4}\,\delta\vf''_{\nu_5 \cdots \nu_s )}}{\delta\vf_{\mu_1 \cdots \mu_s }} + \cdots \right\rbrack G^{\nu_1 \cdots \nu_s}[\phi(\vf)] \nn \\ &=& G^{\mu_1 \cdots \mu_s}[\phi(\vf)]  \eea
 
 \no where the vanishing of multiple traces of the Einstein-like tensor  $G^{[k]}[\phi(\vf)]=0, \, k\ge 2 $ has been used since it follows from $\phi''(\vf)=0$. Therefore the equations of motion stemming from $S_I$ correspond to $G=0$ which on its turn is equivalent to the Fronsdal equation
 for the composite field $\phi(\vf)$,
 
 \be {\cal F}[\phi(\vf)] = 0 \quad . \label{eom_si} \ee
 
 Regarding the gauge symmetries of $S_I$, besides (\ref{weylt}), we also have invariance under CDiff:

\be \dlb  \, \vf = \p \, \bar{\Lambda} \quad . \label{diffbar} \ee 

\no This follows from the fact that  $\dlb \, \vf^{[k]} = 0 $ for any $k\ge 2$, leading to  $\dlb \, \phi = \dlb \, \vf = \p \, \bar{\Lambda} $ which is a known symmetry of the Fronsdal action. Now we are ready to establish the physical equivalence of $S_I$ with the original Fronsdal action. Namely, due to the fact that the double trace $\vf''$ is pure gauge under the Weyl transformation (\ref{weylt}), we can  fix the gauge $\vf''=0$ at action level, see \cite{moto1},  which leads to $S_I[\vf]=S_F[\phi (\vf)]= S_F[\vf]$, whose equations of motion become simply the original Fronsdal equations for a doubly traceless field ${\cal F}[\vf] = 0$.

As a final step toward our goal, we can lift the traceless constraint on the Diff invariance of $S_I$ with the help of the rank-$(s-3)$ compensator field $\alpha$ of subsection 2.1, defining the new action:
 
  \be S_{\alpha}[\vf,\alpha] = S_{F}\left\lbrack \phi[\vft(\vf,\alpha)]\right\rbrack  = \int \, d^Dx\,\, \left\lbrace \frac{\phi(\vft)}2 \, \left[ \cf [\phi(\vft)] - \frac 12 \eta\, {\cal F'}[\phi(\vft)] \right]  \right\rbrace  \label{salpha}\ee

 \no where $\phi(\vft)$ is defined in (\ref{phi1}) with $\vf \to  \vft(\vf,\alpha)$ and $\vft(\vf,\alpha)$  given in (\ref{fredefs}) with $f(\alpha)$ defined in (\ref{fs-1}) with $\Lambda'\to \alpha$. 
 
 From (\ref{diffs}) we see that because of the trace compensation of $\alpha$, although $\vf$ transforms in an unconstrained way, the combination $\vft(\vf,\alpha)$ transforms only under the traceless piece of the gauge parameter which leads to $\dl \vft^{[k]} =0$ with $k\ge 2$. Thus, $\dl \phi[\vft(\vf,\alpha)]= \dl \, \vft = \p \, \bar{\Lambda} $. Since this is a symmetry of the usual Fronsdal action we conclude that indeed we have an unconstrained Diff symmetry,  $\dl S_{\alpha} = 0$,  where  both $\vf$ and $\alpha$ are supposed to transform according to (\ref{diffs}). Moreover, by construction, we still have the rank-$(s-4)$ Weyl invariance (\ref{weylt}) with $\delta_{\lambda} \, \alpha = 0$. 
 
 Concerning the equations of motion $\delta S_{\alpha}/\delta\vf =0$, since 
 $S_{\alpha}$ depends upon $\vf $  only through $\phi[\vft(\vf,\alpha)]$, the same arguments which have led us to (\ref{eom_si}) hold true again. We conclude that $\delta S_{\alpha}/\delta\vf =0$ is equivalent to:

 \be {\cal F}\left\lbrack\phi[\vft(\vf,\alpha)]\right\rbrack = 0 \quad . \label{eom_salpha1} \ee
 
  Since $\vft''$  is invariant under unconstrained Diff and behave  under Weyl transformations (\ref{weylt}) as pure gauge, we can fix  at action level the following gauge without losing physical information while keeping Diff invariance intact,
 
 \be \vft'' = \vf'' -4\, \p \cdot \, \alpha - \p\, \alpha' = 0  \quad . \label{gc2} \ee 
 
 \no In the above gauge we have $\vft^{[k]}=0 $ for $k\ge 2$ and therefore 
 $ \phi[\vft(\vf,\alpha)]= \vft(\vf,\alpha) = \vf - \p\, f(\alpha) $. Consequently, using (\ref{id1}), the unconstrained Diff invariant
  equations of motion
 (\ref{eom_salpha1}) become\footnote{Recall that due to the Stueckelberg nature of the $\alpha$ field, as mentioned before, we do not need to care about the equations of motion $\delta\, S_{\alpha}/\delta\, \alpha = 0 $ which follow trivially from $\delta S_{\alpha}/\delta\vf =0$ and the Diff symmetry.} 
 
 \be  {\cal F}\left[\vf - \p\, f (\alpha )\right] = 
{\cal F}(\vf) - 3\, \p\, \p\, \p\, \alpha =0.  \quad  \label{eom_salpha} \ee 

\no As far as we know, except for the $s=3$ case \cite{schwinger},  the equations (\ref{eom_salpha}) have first appeared in \cite{fscqg} after a truncation of the so called triplet equations obtained from the bosonic string field theory in the tensionless limit. The truncation reduces the spectrum of massless spin $s,s-2,s-4, \cdots, 0$ or $1$, according to even or odd values of $s$  respectively, to only the top spin-s case. The equation (\ref{gc2}) also appears in \cite{fscqg} via the mentioned truncation. In fact, differently from what we have done here, it is possible to deduce (\ref{gc2}) from (\ref{eom_salpha}) by working out a Bianchi identity for ${\cal F}[\vft]$ and assuming vanishing fields at infinity. Here we have shown that this is not surprising since (\ref{gc2}) is the gauge condition behind (\ref{eom_salpha}). We can understand (\ref{eom_salpha}) as the Lagrangian equations of  motion (\ref{eom_salpha1}) stemming from $S_{\alpha}$ after the partial gauge fixing (\ref{gc2}) where Diff plus the rank-$(s-4)$ Weyl symmetry is broken down to simply Diff. In section-4 we show for $s=4$ that this partial gauge fixing from W$_S$Diff down to Diff also works at action level. Finally, notice that we can further use the rank-$(s-3)$ trace $\Lambda'$  of the Diff parameter to turn Diff into traceless Diff via another partial gauge fixing: $\alpha =0$. In this case we go back to the usual Fronsdal equations of motion,
see (\ref{gc2}) and (\ref{eom_salpha}).

	\section{$S_{\alpha}$ and non local actions for $s=4$}
	
	\subsection{ From $S_{\alpha}$ to non local actions}

	Let us now examine the specific case of  $S_{\alpha}$  for spin $s=4$, which can be derived from the general expression (\ref{salpha}) in $D$ spacetime dimensions. The explicit form is given by:
	
	\bea S_{\alpha}[\vf,\alpha] &=&\int d^Dx\,\,\left\lbrack \frac{1}{2} \vf \, \Box \, \vf +2\, (\p\cdot \vf)^2 + 6\, \vf' \, \p\cdot \p \cdot \, \vf - 3\, \vf' \, \Box \, \vf' +3\, (\p \cdot \vf')^2 \nn\right.\\  &+&\left.\frac{3(1+2D)}{2D(D+2)}\,\,\varphi''\,\Box\, \varphi''- 18\,\, \alpha\, \Box\, \p \cdot \vf' -\frac{12(1+2D)}{D(D+2)} \,\,\p\cdot\alpha\, \Box\, \vf'' \nn\right.\\  &+&\left. 12 \,\,\alpha \, \p\cdot \p \cdot \p \cdot \, \vf + 6 \,\p\cdot\alpha \,\,  \p \cdot \p \cdot \, \vf'+ 9\, \alpha\, \Box ^2 \,  \alpha - \frac{3(D-2)(4+5D)}{D\, (D+2)}\,\, \p \cdot \alpha\, \Box  \,\p \cdot \alpha \right\rbrack.\label{lwdiffs4}   \nn\\\eea
	
	\no It is invariant under W$_S$Diff, i.e.,  unconstrained Diff and the Weyl transformation (\ref{weylt}) which becomes for $s=4$ a scalar Weyl, i.e., 
	
	\bea \delta \alpha_{\mu} &=& \Lambda_{\mu}' \nn\\ 	
	\delta \vf_{\mu\nu\alpha\beta} &=& \p_{(\mu}\, \Lambda_{\nu\alpha\beta)} + \eta_{(\mu\nu}\eta_{\alpha\beta)} \, \lambda \quad . \label{ws} \eea

	\no After performing a functional integration over the compensator field $\alpha$, or equivalently, eliminating it through its equations of motion, one obtains the following non-local action still invariant under (\ref{ws}),

	\bea S_{NL}^{W_S}[\vf] &=&\int d^Dx\,\,\left\lbrack \frac{1}{2} \vf \, \Box \, \vf +2\, (\p\cdot \vf)^2 + 6\, \vf' \, \p\cdot \p \cdot \, \vf - 3\, \vf' \, \Box \, \vf' -6\, (\p \cdot \vf')^2 \nn\right.\\ 
	&-&\left.\frac{3(2D+1)}{(D-1)(D+1)}\,\left(2\vf''\,\p\cdot\p\cdot\, \vf'-\frac{1}{2}\vf''\,\Box\, \vf'' -\frac{\vf''\,\p\cdot\p\cdot\p\cdot\p\cdot\,\vf}{\Box}\right) \nn\right.\\ 
	&+&\left. 12\,\frac{\p\cdot\vf'\,\p\cdot\p\cdot\p\cdot\,\vf}{\Box}-4\frac{(\p\cdot\p\cdot\p\cdot\vf)^2}{\Box^2}  \nn\right.\\ 
	&-&\left.\frac{3(D^2-4D-3)}{(D-1)(D+1)}\,\frac{(\p\cdot\p\cdot\,\vf')^2}{\Box}+\frac{6(D^2-2D-2)}{(D-1)(D+1)}\,\frac{\p\cdot\p\cdot\p\cdot\p\cdot\,\vf\,\p\cdot\p\cdot\vf'}{\Box^2}\nn\right.\\ 
	&-&\left. \frac{(D-2)(5D+4)}{2(D-1)(D+1)}\,\frac{(\p\cdot\p\cdot\p\cdot\p\cdot\,\vf)^2}{\Box^3}\right\rbrack.\nn\\ \label{nlwdiff} \eea
	
  Alternatively, we could have first fixed the gauge (\ref{gc2}), which breaks W$_S$Diff down to Diff, and only then integrated over the compensator $\alpha$. Indeed, any vector can be decomposed into its transverse ($\alpha^T$) and longitudinal components, the latter is obtained from (\ref{gc2}), for $s=4$ we have $\p \cdot \alpha = \vf''/4$. Thus, see (\ref{thetaomega}) for the definition of $(\theta,\omega)$,
	
	\be \alpha_{\mu} = \theta_{\mu\nu}\alpha^{\nu} + \omega_{\mu\nu}\alpha^{\nu}  \equiv \alpha_{\mu}^T + \frac{\p_{\mu}(\p \cdot \alpha)}{\Box} = \alpha_{\mu}^T + \frac{\p_{\mu}\, \vf''}{4\, \Box} \quad . \label{alphat} \ee
	
\no After substituting (\ref{alphat}) in (\ref{lwdiffs4}) and integrating over $\alpha^T$, recalling that $\alpha^T$ can only couple to transverse vectors, we end up exactly with the same Diff invariant non local spin-4 action obtained in \cite{fms} after the elimination of $\alpha$ and $\beta$ from the minimal model of \cite{fs2}, see also \cite{francia_int}.  Explicitly we have: 

 \bea {\cal L}_{NL}^{Diff}&=& \frac{1}{2}\vf\, \Box\, \vf +2\, (\p\cdot \vf)^2 + 6\, \vf'\p\cdot\p\cdot \vf- 3\,\vf'\,\Box\,\vf' -6\,(\p\cdot \vf')^2\nn\\ 
 &+& 6\, \vf''\p\cdot\p\cdot\vf'- \frac{3\,\vf''\p\cdot\p\cdot\p\cdot\p\cdot\vf}{\Box}-\frac{3}{2}\,\vf''\,\Box\,\vf''+\frac{12\, \p\cdot\vf'\p\cdot\p\cdot\p\cdot\vf}{\Box}\nn\\
 &-&\frac{9\,(\p\cdot\p\cdot\vf')^2}{\Box}-\frac{4\,(\p\cdot\p\cdot\p\cdot\vf)^2}{\Box^2}+\frac{12\,\p\cdot\p\cdot\p\cdot\p\cdot\vf\, \p\cdot\p\cdot\vf'}{\Box^2}-\frac{4\,(\p\cdot\p\cdot\p\cdot\p\cdot\vf)^2}{\Box^3}\nn\\ \label{diffs4} \eea 
	
	\no Instead of  examining the specific particle content of (\ref{nlwdiff}) and (\ref{diffs4}), we consider in the next subsection a broader class of Diff invariant non local actions depending on two arbitrary real constants $(a,b)$. Such analysis will emphasize the  uniqueness of  (\ref{nlwdiff}) and (\ref{diffs4}) within their symmetry classes.
	
\subsection{Diff invariant non local spin-4 actions} 

Differently from the $s=2$ case it is known that for $s\ge 3$  there is no local Lagrangian, in terms of only one rank-$s$ symmetric field, invariant under unconstrained Diff. Giving up locality we are led to a two parameters action invariant under $\dl \vf_{\mu_1 \cdots \mu_s} = \p_{(\mu_1}\, \Lambda_{\mu_2 \cdots \mu_{s} )}$,

\be S_{(a,b)}=\frac 12 \, \int d^Dx\left\lbrack \vf \,\Box\left( P_{11}^{(4)}+b\,P_{11}^{(2)}+a\,P_{11}^{(0)}\right)\,\vf. \right\rbrack \label{sab2}  \ee

\no where the spin-s projection operators $P_{IJ}^{(s)}$ acting on rank-4 symmetric fields are given in Appendix A. More explicitly we have,

	\bea S_{(a,b)}&=&\int d^Dx\left\lbrack \frac{1}{2} \vf \, \Box \, \vf +2\, (\p\cdot \vf)^2 +\frac{ 3 \, (\p\cdot \p \cdot \, \vf)^2}{\Box} + \frac{2\, (\p\cdot \p \cdot \p \cdot \, \vf)^2}{\Box^2}+\frac{\, (\p\cdot \p \cdot \p \cdot\p \cdot \, \vf)^2}{2\Box^3}\nn\right.\\ 
	&+&\left.c_1(b)\left(\vf' \, \Box \, \vf' +2\,(\p\cdot \vf')^2 -2\,\vf'\p\cdot\p\cdot \vf +\frac{ \, (\p\cdot \p \cdot \, \vf')^2}{\Box}-\frac{4\,\p\cdot \vf'\,\p\cdot \p \cdot \p \cdot \, \vf}{\Box} \nn\right.\right.\\ 
	&+&\left.\left.\frac{(\p\cdot\p\cdot\,\vf)^2}{\Box}-\frac{2\,\p\cdot\p\cdot \vf'\,\p\cdot \p \cdot \p \cdot\p\cdot \, \vf}{\Box^2} + \frac{2\, (\p\cdot \p \cdot \p \cdot \, \vf)^2}{\Box^2}+\frac{\, (\p\cdot \p \cdot \p \cdot\p \cdot \, \vf)^2}{\Box^3}\right)\nn\right.\\ 
	&+&\left.c_2(a,b)\left(\vf'' \, \Box \, \vf'' -4\,\vf''\p\cdot\p\cdot \vf'+\frac{4\,(\p\cdot\p\cdot\vf')^2}{\Box}+\frac{2\,\vf''\p\cdot\p\cdot\p\cdot\p\cdot\vf}{\Box}\right.\right.\nn\\
	&-&\left.\left. \frac{4\,\p\cdot\p\cdot \vf'\p\cdot\p\cdot\p\cdot\p\cdot\vf}{\Box^2}+\frac{(\p\cdot\p\cdot\p\cdot\p\cdot\vf)^2}{\Box^3} \right)\right\rbrack.\nn\\ \label{nlab} \eea
	\no In order to improve the readability of the expression, we  introduce the coefficients $c_1$ and $c_2$:

	\be c_1(b)= \frac{3(b-1)}{(D+3)}\quad; \quad c_2(a,b)=\frac{3\left\lbrack (D-1)-2b(D+1)+a(D+3)\right\rbrack}{2\,(D-1)(D+1)(D+3)}.\ee

	In what follows  we will analyse the unitarity conditions on (\ref{nlab})  . After the introduction of a symmetric rank-4 source $J_{\mu\nu\alpha\beta}$ subjected to appropriate constraints dictated by the corresponding gauge symmetry, the general form of the Lagrangian density in (\ref{nlab})   (suppressing indices) is given by:
	 
	\be {\cal L} (a,b) = \frac 12 \vf \,  G(\p) \, \vf + \vf \, J \quad . \label{lab} \ee 
	
	\no The kinetic non local differential operator $G(\p)$ depends on $(a,b)$ and is read off from (\ref{nlab}). Regarding the particle content, the key quantity is the two-point amplitude. It is given basically by the propagator saturated with the sources. In the momentum space,  suppressing indices again, we have
	
	\be {\cal A}_2 (k) = -\frac{i}{2} J^* (k)G^{-1}(\p \to i\, k) J(k) \quad. \label{a2k} \ee 
	
\no If the imaginary part of the residue at the massless pole  $(k^{\mu}k_{\mu} =0)$  of $ {\cal A}_2 (k)$ is positive we have a physical  particle, otherwise we have a ghost. In order to obtain $G^{-1}(\p)$ we have introduced   in Appendix A a basis  of non local differential operators acting on rank-4 symmetric fields. The analysis will be divided into four cases as shown in table 1. The actions (\ref{diffs4}) and (\ref{nlwdiff}) belong to the cases I and II respectively.


\begin{table}
		\centering
		\begin{tabular}{l|l|l} 
			\hline
			{ Cases }	& { $(a,b)$}  & { Symmetry }   \\
			\hline
			\hline
			Case I \, (Diff)	& $a \ne 0$ and $b \ne 0$                                                                         & $\delta\, \vf_{\mu\nu\alpha\beta} = \p_{(\mu}\, \Lambda_{\nu\alpha\beta)} $\\
			Case II \, (W$_S$Diff)	& $a = 0$ and $b \ne 0$                                                                         & $\delta\,  \vf_{\mu\nu\alpha\beta} = \p_{(\mu}\, \Lambda_{\nu\alpha\beta)}   + \eta_{(\mu\nu} \, \eta_{\alpha\beta)} \, \lambda  $\\
			Case III \, (W$_{TT}$Diff)	& $a \ne 0$ and $b = 0$                                                                         & $\delta\,  \vf_{\mu\nu\alpha\beta} = \p_{(\mu}\, \Lambda_{\nu\alpha\beta)}   +\eta_{(\mu\nu}\bar{\lambda}^T_{\alpha\beta)} $\\
			Case IV \, (WDiff)	& $(a,b)=(0,0)$                                                                         & $\delta\,  \vf_{\mu\nu\alpha\beta} = \p_{(\mu}\, \Lambda_{\nu\alpha\beta)}  +\eta_{(\mu\nu}\lambda_{\alpha\beta)}  $\\
			
			\hline
		\end{tabular}
		\caption{Constraints on $(a,b)$  and  corresponding symmetries. Note that $\p^{\alpha}\bar{\lambda}^T_{\alpha\beta}=0=\eta^{\alpha\beta}\bar{\lambda}^T_{\alpha\beta}$.}
	\end{table}

 \subsubsection{Case I (Diff)}

	We add to (\ref{lab}) the following gauge-fixing term in order to obtain $G^{-1}(\p)$:
	
	\be {\cal L}_{g.f}^{(1)}= \frac{1}{\lambda_1}(\p\cdot \vf)^2.\label{gf1}\ee
	
	\no where $\lambda_1$ is an arbitrary gauge fixing parameter (from now on, the gauge fixing terms will always be introduced with a gauge fixing parameter $\lambda_i$ with $i=1,2,3$ and 4). So we derive the following inverse kinetic operator (suppressing indices):

	\be G^{-1}=\frac{2\, P_{11}^{(4)}}{\Box}-\frac{4\lambda_1 P_{11}^{(3)}}{\Box}+\frac{2\, P_{11}^{(2)}}{b\,\Box}-\frac{2\lambda_1 P_{22}^{(2)}}{\Box}-\frac{4\lambda_1 P_{11}^{(1)}}{\Box}-\frac{4\lambda_1 P_{22}^{(1)}}{3\Box}+\frac{2\,P_{11}^{(0)}}{a\,\Box}-\frac{2\lambda_1 P_{22}^{(0)}}{\Box}-\frac{\lambda_1 P_{33}^{(0)}}{\Box} \\ \label{propa1} \ee
	
	\no Using a totally symmetric conserved source $J^{\mu\nu\alpha\beta}$ in order to preserve the Diff symmetry of the action, after some technical steps, one obtains from (\ref{a2k}) and (\ref{propa1}) the following expression:
	\be {\cal A}_2(k)=\frac{i}{k^2}\left\lbrace J^*_{\mu\nu\alpha\beta}J^{\mu\nu\alpha\beta}-\frac{6(b-1)}{(D+3)b}J^*_{\mu\nu}J^{\mu\nu}+\frac{3\left\lbrack (D-1) ab-2(D+1)a+(D+3)b\right\rbrack \vert J\vert^2}{ab(D-1)(D+1)(D+3)}\right\rbrace.\ee
	
	\no where $J_{\mu\nu}=\eta^{\alpha\beta}J_{\mu\nu\alpha\beta}$ and $J=\eta^{\mu\nu}J_{\mu\nu}$.  Note the presence of singularities at $ab=0$ indicating the appearance of extra symmetries at $ab=0$. 
	
	Because of the Lorentz invariance of ${\cal A}_2(k)$, once we have cancelled the massless pole when calculating the residue: $R_0=\lim_{k^2\to 0} k^2 \, {\cal A}_2(k)$, we may choose the convenient frame $k_{\mu}=(k_0,0,0,\cdots,0,k_0)$ which altogether with the restriction on the source $k_{\mu}J^{\mu\nu\lambda\alpha} = 0$  leads to the following expression for the residue:

	\bea R_0 &=&  i\, \left\lbrace \vert \bar{J}_{ijkl}\vert^2+\frac{6(b+(D+2))}{(D+2)(D+3)b}\vert \bar{J}_{ij}\vert^2\right.\nn\\
	&+&\left.\frac{3\left\lbrack 3(D-1) ab+2D(D+1)a+D(D-2)(D+3)b\right\rbrack \vert J\vert^2}{ab(D-1)(D+1)(D+3)(D-2)}\right\rbrace.\label{Adiff}\eea
	
	\no where ${i,j,k,l = 1,2,...,D-2}$. The bar indicates a spatially traceless quantity, explicitly,
	
	\bea J_{ijkl}&=&\bar{J}_{ijkl}+\frac{1}{(D+2)}\delta_{(ij}\bar{J}_{kl)}+\frac{1}{D(D-2)}\delta_{(ij}\delta_{kl)}J\eea
	
	\be J_{ij}=\bar{J}_{ij}+\frac{1}{D-2}\delta_{ij}J\quad;\quad \delta_{kl}J_{ijkl}=J_{ij}\ee

\be  \delta_{ij}\bar{J}_{ijkl}=0\quad;\quad \delta_{ij}\bar{J}_{ij}=0\quad;\quad J=\delta_{ij}\delta_{kl}J_{ijkl}\quad;\quad \delta_{ij}\delta_{ij}=(D-2)\ee

\no Regarding the number of independent components of the tensors, the reader can check that,
\bea N(\bar{J}_{ijkl})&=& \frac{(D-2)(D-1)(D-3)(D+4)}{4!} \quad  \,  \label{N1} \\
 N(\bar{J}_{ij})&=&  \frac{D(D-3)}{2} \quad . \label{N2}\eea

	\no Notice that $N(\bar{J}_{ijkl})=2= N(\bar{J}_{ij})$ at $D=4$. This is the right number of independent helicities ($\pm s$) for a massless particle of arbitrary integer spin-$s$ with $s>0$, including the respective cases of $s=4$ and $s=2$. From (\ref{Adiff}), one can infere the particle content displayed in Table 2. Note that the non-local action obtained in \cite{fms}, see ({\ref{diffs4}), corresponds to the case $b = -(D+2)$ and $a = -D(D+2)$. This is the only case where we have the propagation of only spin-4 massless modes as in the Fronsdal theory, so ({\ref{diffs4}) is the unique  non local  action describing massless spin-4 particles which is invariant only under unconstrained Diff, thus confirming the analysis of \cite{fms}. 
	
	Since the technical steps in the next three cases are essentially the same ones, henceforth we only mention the different gauge fixing conditions  and suppress the formulas for the propagators by going directly to the amplitude.

	\begin{table}
		\centering
		\begin{tabular}{lll} 
			\hline
			{  $b$}	& {  $a$}  & { Particle Content}   \\
			\hline
			\hline
			$b=-(D+2)$	& $a=-D(D+2)$                                                                        & $spin: 4$   \\
			$b<-(D+2)\quad or \quad b>0$	& $a=-\frac{ b D(D-2)(D+3)}{3 b(D-1)+2D(D+1)}$                   & $spins: 4\,,\,2$  \\
			$b<-(D+2)\quad or \quad b>0$	    & $a<-\frac{ b(D-2)D(D+3)}{3 b(D-1)+2D(D+1)}\quad or\quad a>0$                                        & $spins: 4\,,\, 2\,, 0$  \\
			$b=-(D+2)$	& $a<-D(D+2)\quad or \quad a>0$                                                      & $spins: 4\,,\, 0$  \\
			
			\hline
		\end{tabular}
		\caption{Unitarity conditions	    on $(a,b)$ and the particle content of the Diff invariant action (\ref{diffs4}) with $ab\ne 0$. Note that $3 b(D-1)+2D(D+1) \ne 0$ if $b<-(D+2)\quad or \quad b>0$. }
	\end{table}

 \subsubsection{Case II (W$_S$Diff)}

	In this subsection we analyze the particle content of the W$_S$Diff  invariant model, i.e., when $a=0$ and $b\ne 0$. With this choice, the action $S(0,b)$ is invariant under (\ref{ws}). In terms of the spin projection operators the lagrangian density becomes: 
	\be {\cal L}=\frac 12 \,\vf \,\Box\left( P_{11}^{(4)}+b\,P_{11}^{(2)}\right)\,\vf, \ee
	
\no and	we need to consider an extra gauge-fixing term, in order to fix the Weyl symmetry, in addition to the one given by (\ref{gf1}). A natural choice is:
	
	\be {\cal L}_{g.f}^{(2)}=\frac{(\vf'')^2}{\lambda_2}.\ee
	
	\no After saturating with the constrained sources we get rid of the arbitrary constants $(\lambda_1,\lambda_2)$ and end up with the amplitude:
	
	\be {\cal A}_2(k)=\frac{i}{k^2}\left\lbrace J^*_{\mu\nu\alpha\beta}J^{\mu\nu\alpha\beta}-\frac{6(b-1)}{(D+3)b}J^*_{\mu\nu}J^{\mu\nu}\right\rbrace.\ee
	\no We have used a conserved $(k_{\mu}J^{\mu\nu\alpha\beta} = 0)$ and double traceless $(J'' = 0)$ source. This leads, in the same frame used in case I, to the following expression for the residue at the massless pole:
	\bea R_0 = i \, \left\lbrack \vert \bar{J}_{ijkl}\vert^2+\frac{6\,[b+(D+2)]}{(D+2)(D+3)b}\vert \bar{J}_{ij}\vert^2\right\rbrack.\label{Adiff2}\eea
	
	\no  The particle content of the model is organized in Table 3. Only in the case $b=-(D+2)$, which corresponds to our 
	W$_S$Diff non local action (\ref{nlwdiff}), we have only spin-4 massless modes which makes (\ref{nlwdiff}) the unique non local massless spin-4 model with only W$_S$Diff symmetry.

	For completeness, in the next two cases we analyze the particle content of models invariant under alternative Weyl plus Diffeomorphism transformations, which we call W$_{TT}$Diff and WDiff since we have a rank-2 gauge parameter which is traceless and transverse (TT) in the case III while fully unrestricted in the case IV. 
	
	\begin{table}
		\centering
		\begin{tabular}{lll} 
			\hline
			{  $b$}	  & { Particle Content}   \\
			\hline
			\hline
			$b=-(D+2)$	 & $spin: 4$   \\
			$b<-(D+2)\quad or \quad b>0$	 & $spins: 4\,,\,2$  \\
			
			\hline
		\end{tabular}
		\caption{Unitarity conditions on $b$ and the particle content of the case II (W$_S$Diff), $a=0$.}
	\end{table}

\subsubsection{ Case III (W$_{TT}$Diff) }

Now we consider $a\ne 0 $ and $b=0$, which implies that the lagrangian density is given by:
\be {\cal L}=\frac 12 \,\vf \,\Box\left( P_{11}^{(4)}+ a\,P_{11}^{(0)}\right)\,\vf. \ee

\no One can verify that the action is invariant under unconstrained Diff and a Weyl type transformation with a traceless and transverse rank-2 tensor (W$_{TT}$Diff), namely:

\be \delta\vf_{\mu\nu\alpha\beta}=\p_{(\mu}\Lambda_{\nu\alpha\beta)}+\eta_{(\mu\nu}\bar{\psi}^T_{\alpha\beta)},\ee

\no where $\eta^{\alpha\beta}\bar{\psi}^T_{\alpha\beta}=0=\p^{\alpha}\bar{\psi}^T_{\alpha\beta)}$. In addition to the gauge fixing term (\ref{gf1}) we choose the following gauge condition in terms of a traceless and  transverse tensor:

\be {\cal L}_{g.f}^{(3)}=\frac{1}{\lambda_3}\left\lbrack \Box^2 \vf'_{\mu\nu}-\Box(\p_{\mu}\p^{\alpha}\vf'_{\alpha\nu}+\p_{\nu}\p^{\alpha}\vf'_{\alpha\mu})+\p_{\mu}\p_{\nu}\p\cdot\p\cdot\vf'-\frac{\Box\theta_{\mu\nu}}{(D-1)}(\Box \vf''-\p\cdot\p\cdot \vf') \right\rbrack^2.\ee

Noticing that the W$_{TT}$Diff symmetry requires that $k^{\alpha}J_{\mu\nu\alpha\beta}=0$ and $J_{\mu\nu}=\theta_{\mu\nu}J/(D-1)$. Consequently, after obtaining the propagator one can check that the amplitude is given by:
\be {\cal A}(k)=\frac{1}{k^2}\left\lbrack J^*_{\mu\nu\alpha\beta}J^{\mu\nu\alpha\beta}+\frac{3(1-a)}{a(D+1)(D-1)}\vert J\vert ^2\right\rbrack,\ee

\no and the residue, in the same previously used reference frame, becomes:

\be R_0 = i\, \left\lbrack \vert \bar{J}_{ijkl}\vert^2+\frac{3(a+D(D+2))\vert J\vert^2}{aD(D+1)(D+2)(D-1)}\right\rbrack \label{ro3} \ee

\no In (\ref{ro3}) we have  $J=\delta^{ij}\delta^{kl}J_{ijkl}$ which turns out to be equal $\eta^{\mu\nu}\eta^{\alpha\beta}J_{\mu\nu\alpha\beta}$ in the frame we are working on.  The particle content one can derive from  (\ref{ro3}) is displayed in Table 4.	
	
	\begin{table}
	\centering
	\begin{tabular}{lll} 
		\hline
		{ $a$}	  & { Particle Content}   \\
		\hline
		\hline
		$a=-D(D+2)$	 & $spin: 4$   \\
		$a<-D(D+2)\quad or \quad a>0$	 & $spins: 4\,,\,0$  \\
		
		\hline
	\end{tabular}
	\caption{Unitarity conditions on $a$ and the particle content of the case III (W$_{TT}$Diff),  $b=0$ .}
\end{table}

\subsubsection{ Case IV (WDiff) }

Our last case corresponds to $a=b=0$ which is given by the quite simple expression:
\be {\cal L}=\frac 12 \,\vf \,\Box\left( P_{11}^{(4)}\right)\,\vf, \ee
such lagrangian density becomes invariant under what we are calling WDiff symmetry, given by:
\be \delta\vf_{\mu\nu\alpha\beta}=\p_{(\mu}\Lambda_{\nu\alpha\beta)}+\eta_{(\mu\nu}\psi_{\alpha\beta)},\ee
with $\psi_{\alpha\beta}$ an arbitrary symmetric rank-2 tensor. In addition to the gauge fixing term  (\ref{gf1}) we now have to consider an extra one in order to fix the tensorial Weyl part. This is given by:
\be {\cal L}_{g.f}^{(3)}=\frac{(\vf')^2}{\lambda_4}\ee

\no In this case the source is subjected to the following restrictions $k^{\mu}J_{\mu\nu\alpha\beta}=0$ and $J'=0$. One can verify that with these conditions the transition amplitude becomes simply:
\be {\cal A}_2(k)= i\, \frac{J^*_{\mu\nu\alpha\beta}J^{\mu\nu\alpha\beta}}{k^2}\ee
then, after considering the conditions on the source the residue becomes
 \be R_0 =  i\, \vert \bar{J}_{ijkl}\vert^2, \label{r04}\ee   
and it is straightforward to conclude that the model propagates only a spin-4 particle\footnote{From (\ref{N1}) and (\ref{N2}) we recall that there is no spin-4 nor spin-2 propagation in $D=3$. So the particle content in the case IV and in tables 2,3 and 4 hold for $D>3$. Only the spin-0 particle in tables 2 and 4 survives in $D=3$}.	It is the spin-4 analogue of a nonlocal  linearized conformal gravity. We believe that it can be made local via the introduction of a rank-2 symmetric compensator.

	\section{Conclusions}
	
	Here we have obtained a local action $S_{\alpha}[\vf,\alpha]$, see (\ref{salpha}), in terms of only two symmetric fields of rank-$s$ $(\vf)$ and  $(s-3)$ $(\alpha)$ describing free massless spin-$s$ particles, on flat Minkowski space, for arbitrary integer $s$. There is no double traceless condition on the field $\vf$ nor any constraint  on the higher spin (HS) diffeomorphism (Diff) parameter due to the compensating field $\alpha$\footnote{ In the recent work \cite{hp} several actions for irreducible and reducible higher spins have been investigated with the help of compensating fields. Differently from us they start with traceless fields.}. The action $S_{\alpha}$ is invariant under W$_S$Diff which corresponds to a Weyl like rank $(s-4)$ symmetry, see (\ref{weylt}),
	altogether with unconstrained HS diffeomorphisms ($\delta \vf = \p\, \Lambda $) where $\Lambda$ is a totally symmetric rank-$(s-1)$ gauge parameter without any further restriction. 
	
	The action $S_{\alpha}$ is obtained from the Fronsdal action via a field redefinition $\phi \to \phi[\vft(\vf,\alpha)]$ defined in (\ref{fredefs}) and (\ref{phi1}). Its equations of motion, after the partial gauge fixing condition (\ref{gc2}), become exactly  the equations stemming from the truncation of the tensionless limit of the open string field theory spectrum to only spin-$s$ particles described by symmetric tensors, see (\ref{eom_salpha}). Those equations also appear in the ``minimal'' model $S_{FS}[\vf,\alpha, \beta]$ of \cite{fs2}, see also \cite{fms},  after the elimination of the rank-$(s-4)$ field $\beta$.
	
	 In the specific $s=4$ case we have shown in section 4 that the connection between $S_{\alpha}$ and $S_{FS}$ goes beyond on-shell, since the nonlocal Diff invariant action written in terms only of the physical field $\vf$, obtained after integration over $\alpha$ and $\beta$, see also  \cite{francia_int}, can be reproduced starting from $S_{\alpha}$ and integrating over $\alpha$ in the gauge ({\ref{gc2}) which breaks W$_S$Diff down to Diff. We conjecture that this is true for arbitrary spin-$s$. This touches a key point in the comparison between the W$_S$Diff action $S_{\alpha}[\vf,\alpha]$ and the Diff action $S_{FS}[\vf,\alpha, \beta]$. Namely, in $S_{FS}$ the double traceless condition on both the field $\vf$ and the source $J$, which just need to be conserved  $\p \cdot J =0$, is lifted while in our model the source must be not only conserved but also double traceless due to the Weyl symmetry. In this sense our model is closer to the original Fronsdal theory, however as we have shown explicitly for $s=4$, the nonlocal Diff invariant model  ${\cal L}_{NL}^{Diff}$, for which no double traceless condition on the source is required, can also be deduced from $S_{\alpha}$ after integrating over $\alpha$. 
	 
	In  subsection 4.2  we have introduced a broader class of non local actions for rank-4 symmetric fields parameterized by two real constants $(a,b)$, as shown in (\ref{nlab}). With the aid of the spin operators presented in Appendix A, we have deduced unitarity conditions for four different cases. The first two cases  have shown how special are the actions  (\ref{nlwdiff}) and (\ref{diffs4}) obtained from $S_{\alpha}$ since they both correspond to the unique action propagating only spin-4 modes within their respective symmetry class W$_S$Diff and Diff. For completeness we have also investigated other two cases with traceless and transverse (W$_{TT}$Diff)  and unrestricted (WDiff) Weyl parameter. The results are summarized in tables 1-5. 	 	
	 	
	 		In summary,  the original Fronsdal theory corresponds to a partial gauge fixing ($\alpha=0=\vf''$)  of $S_{\alpha}$ where the WDiff symmetry is reduced to constrained (traceless) Diff while ${\cal L}_{NL}^{Diff}$ comes from another partial gauge fixing where W$_S$Diff reduces do Diff (unconstrained). Regarding a possible HS geometry, we are currently trying to figure out whether there is some W$_S$Diff geometrical interpretation of our equations of motion before gauge fixing. Since we have obtained several non local actions in section 4 as opposed to the great majority of local actions in field theory, it is appropriate to mention that non locality in massless HS theories is not totally surprisingly. From a possible geometrical point of view regarding the linearizecd HS gauge symmetries, the linearized HS curvatures suggested in  \cite{wit} contain $s$ derivatives of the rank-s symmetric gauge field. Thus, for dimensional reasons we do expect negative powers of the operator $\Box = \p^{\mu}\p_{\mu}$ in a covariant massless theory after eliminating extra fields.
	
	There are different directions we could follow now. In particular, we are presently investigating the extension of $S_{\alpha}$  to the $AdS$ background. It looks very promising since  $S_{\alpha}$ corresponds to a field redefinition of the Fronsdal theory for which such extension (deformed Fronsdal tensor) has been derived long ago \cite{fronsdal_ads}. The extension to fermions \cite{fang} is  in progress and does not seem to offer any severe obstruction either. It is natural also to extend our approach to mixed symmetry tensors and compare it with \cite{campo_mix1, campo_mix2}, see also the review work \cite{campoleoni_r}, where those fields have been considered. The approach of \cite{fs1} and the  BRST approach have been geralized for the massive case in  \cite{francia_m1,fms_massive,francia_m2} and \cite{bkl_massive, quartet_massive} respectively and we are also currently investigating the massive generalization of our results as well as the introduction of extra fields to bring down our higher derivative model $S_{\alpha}$ to second order as carried out in \cite{francia_m1} where the introduction of three extra auxiliary fields, besides the couple $(\alpha,\beta)$, has brought the minimal model \cite{fs2} to second order in derivatives.. The remark, see \cite{singh} and a note in \cite{francia_m2}, that the Singh-Hagen \cite{shb} theory for massive arbitrary spin-$s$ free particles can be reformulated in terms of only two (unconstrained) fields of rank $s$ and $(s-3)$ makes the search of a massive version of the present work rather natural.
	
	Another possible direction to follow is to apply a similar procedure in the massless Higher Spin model of \cite{sv} where the gauge parameter is traceless and transverse.   In fact, we believe that the  method used here of introducing Stueckelberg fields in massless theories in order to undo constraints on  gauge parameters  may find more general applications in gauge  theories.

	It is worth mentioning that in several steps and sections of the present work we have checked our results with the help of the {\it xTras/xAct Mathematica} package guided mainly by the lines and procedures of \cite{Teake}.

	\section*{Acknowledgements}
	
	 DD is partially supported by CNPq  (grant 313559/2021-0). RSB is supported by CAPES.
	
	\appendix 	
	\section{Rank-4 spin-projection operators}\label{app:A}
	
	Here, we introduce a complete basis of spin projection and transition operators on totally symmetric rank-4 tensors in $D$ dimensions. This result is a byproduct of our work and generalizes the previous work \cite{Mendonca1}, where some of us developed spin-3 projection operators. We believe that this result may also be useful for future updates of the {\it PSALter} package \cite{Barker}, which is currently available for the unitarity analysis of theories involving up to rank-3 tensors. Explicitly we have,

	\begin{eqnarray}
		(P^{(4)}_{11})^{(\mu\nu\rho\sigma)}_{(\alpha\beta\gamma\lambda)} &=&   \theta^{(\mu}_{(\alpha} \theta^\nu_\beta \theta^\rho_\gamma \theta^{\sigma)}_{\lambda)} - (P^{(2)}_{11})^{(\mu\nu\rho\sigma)}_{(\alpha\beta\gamma\lambda)}-(P^{(0)}_{11})^{(\mu\nu\rho\sigma)}_{(\alpha\beta\gamma\lambda)}, \\		
		(P^{(2)}_{11})^{(\mu\nu\rho\sigma)}_{(\alpha\beta\gamma\lambda)} &=&\frac{6}{(D+3)}\,\,\left\lbrack \theta_{(\alpha\beta}\theta^{(\mu\nu}\theta^\rho_\gamma \theta^{\sigma)}_{\lambda)} -\frac{(D+1)}{3}\,\,(P^{(0)}_{11})^{(\mu\nu\rho\sigma)}_{(\alpha\beta\gamma\lambda)}\right\rbrack\\ 
		(P^{(0)}_{11})^{(\mu\nu\rho\sigma)}_{(\alpha\beta\gamma\lambda)} &=& \frac{3}{(D-1)(D+1)} \,\, \theta_{(\mu\nu} \theta_{\gamma\lambda)}  \theta^{(\alpha\beta}\theta^{\rho\sigma)},  \\
		(P^{(3)}_{11})^{(\mu\nu\rho\sigma)}_{(\alpha\beta\gamma\lambda)} &=& 4\,\,  \theta^{(\mu}_{(\alpha} \theta^\nu_\beta \theta^\rho_\gamma \omega^{\sigma)}_{\lambda)}-(P^{(1)}_{11})^{(\mu\nu\rho\sigma)}_{(\alpha\beta\gamma\lambda)},  \\
		(P^{(1)}_{11})^{(\mu\nu\rho\sigma)}_{(\alpha\beta\gamma\lambda)} &=& \frac{12}{(D+1)}\theta_{(\alpha\beta}\theta^{(\mu\nu}\theta^\rho_\gamma \omega^{\sigma)}_{\lambda)} \\
		(P^{(2)}_{22})^{(\mu\nu\rho\sigma)}_{(\alpha\beta\gamma\lambda)} &=& 6\,\,\theta^{(\mu}_{(\alpha} \theta^\nu_\beta \omega^\rho_\gamma \omega^{\sigma)}_{\lambda)}-(P^{(0)}_{22})^{(\mu\nu\rho\sigma)}_{(\alpha\beta\gamma\lambda)},\\
		(P^{(0)}_{22})^{(\mu\nu\rho\sigma)}_{(\alpha\beta\gamma\lambda)} &=& \frac{6}{(D-1)}\,\,\theta^{(\mu}_{(\alpha} \theta^\nu_\beta \omega^\rho_\gamma \omega^{\sigma)}_{\lambda)},\\
		(P^{(1)}_{22})^{(\mu\nu\rho\sigma)}_{(\alpha\beta\gamma\lambda)} &=& 4\,\,\theta^{(\mu}_{(\alpha} \omega^\nu_\beta \omega^\rho_\gamma \omega^{\sigma)}_{\lambda)},\\
		(P^{(0)}_{33})^{(\mu\nu\rho\sigma)}_{(\alpha\beta\gamma\lambda)} &=& \,\,\omega^{\mu}_{\alpha} \omega^\nu_\beta \omega^\rho_\gamma \omega^{\sigma}_{\lambda},. 
	\end{eqnarray}
	
\no where the following projection operators on vectors have been used

\be \omega_{\mu\nu} = \frac{\p_{\mu}\p_{\nu}}\Box \quad ; \quad \theta_{\mu\nu} = \eta_{\mu\nu} - \omega_{\mu\nu}  \quad . 	\label{thetaomega} \ee	
	
	As a requirement, the basis must be orthonormal, and the projectors must be idempotent:
	\begin{eqnarray}
		P_{ij}^{(s)}P_{kl}^{(r)}=\delta^{sr}\delta_{jk}P_{il}^{(s)}.\label{algebra}
	\end{eqnarray}

	In our notation, the superscripts ($r$) and ($s$) denote the spin subspace, while the subscripts $i, j, k, l$ are used to distinguish between projectors and transition operators. Specifically, when $i = j$ or $k = l$, we have a projector; on the other hand, if $i \neq j$ or $k \neq l$, we have a transition operator. Additionally, subscripts are employed to count the number of projectors in a given spin subspace. The set of projectors decomposes the identity as follows:
	
	
	\begin{eqnarray}
		\sum_{i,s} P^{(s)}_{ii}  = \id , \label{id}
	\end{eqnarray}
	where $\id$ stands for the symmetric rank-4 identity operator, i.e.:
	\begin{eqnarray}
	\id^{(\mu\nu\rho\sigma)}_{(\alpha\beta\gamma\lambda)} =  \delta^{(\mu}_{(\alpha} \delta^\nu_\beta \delta^\rho_\gamma \delta^{\sigma)}_{\lambda)}.
	\end{eqnarray}
	Finally, the transition operators $P^{(s)}_{ij}$ for $s=0 , 1$ and 2 are given respectively by:
	\begin{eqnarray}
		(P^{(0)}_{{12}})^{(\mu\nu\rho\sigma)}_{(\alpha\beta\gamma\lambda)} &=& \frac{6}{(D-1)\sqrt{2(D+1)}}   \theta^{(\mu\nu}\omega^{\rho\sigma)}\theta_{(\alpha\beta}\theta_{\gamma\lambda)}  \\ 
		(P^{(0)}_{{21}})^{(\mu\nu\rho\sigma)}_{(\alpha\beta\gamma\lambda)} &=& \frac{6}{(D-1)\sqrt{2(D+1)}}   \theta^{(\mu\nu}\theta^{\rho\sigma)}\theta_{(\alpha\beta}\omega_{\gamma\lambda)},  \\
		(P^{(0)}_{{13}})^{(\mu\nu\rho\sigma)}_{(\alpha\beta\gamma\lambda)} &=&  \frac{3}{\sqrt{3(D-1)(D+1)}}   \omega^{(\mu\nu}\omega^{\rho\sigma)}\theta_{(\alpha\beta}\theta_{\gamma\lambda)},   \\ 
		(P^{(0)}_{{31}})^{(\mu\nu\rho\sigma)}_{(\alpha\beta\gamma\lambda)} &=& \frac{3}{\sqrt{3(D-1)(D+1)}}   \theta^{(\mu\nu}\theta^{\rho\sigma)}\omega_{(\alpha\beta}\omega_{\gamma\lambda)},\\
		(P^{(0)}_{{23}})^{(\mu\nu\rho\sigma)}_{(\alpha\beta\gamma\lambda)} &=& \frac{6}{\sqrt{6(D-1)}}   \omega^{(\mu\nu}\omega^{\rho\sigma)}\theta_{(\alpha\beta}\omega_{\gamma\lambda)},\\
		(P^{(0)}_{{32}})^{(\mu\nu\rho\sigma)}_{(\alpha\beta\gamma\lambda)} &=& \frac{6}{\sqrt{6(D-1)}}   \theta^{(\mu\nu}\omega^{\rho\sigma)}\omega_{(\alpha\beta}\omega_{\gamma\lambda)}\\
		(P^{(1)}_{{12}})^{(\mu\nu\rho\sigma)}_{(\alpha\beta\gamma\lambda)} &=& \frac{12}{\sqrt{3(D+1)}}   \theta_{(\alpha\beta}\theta^{(\mu}_{\gamma}\omega^{\nu}_{\lambda)}\omega^{\rho\sigma)},\\
		(P^{(1)}_{{21}})^{(\mu\nu\rho\sigma)}_{(\alpha\beta\gamma\lambda)} &=& \frac{12}{\sqrt{3(D+1)}}   \theta^{(\mu\nu}\theta^{\rho}_{(\alpha}\omega^{\sigma)}_{\beta}\omega_{\gamma\lambda)}\\
		(P^{(2)}_{{12}})^{(\mu\nu\rho\sigma)}_{(\alpha\beta\gamma\lambda)} &=& \frac{6}{\sqrt{(D+3)}}   \left\lbrack \theta_{(\alpha\beta}\theta^{(\mu}_{\gamma}\theta^{\nu}_{\lambda)}\omega^{\rho\sigma)} -3\sqrt{2(D+1)}( P^{(0)}_{{12}})^{(\mu\nu\rho\sigma)}_{(\alpha\beta\gamma\lambda)}\right\rbrack,\\
		(P^{(2)}_{{21}})^{(\mu\nu\rho\sigma)}_{(\alpha\beta\gamma\lambda)} &=& \frac{6}{\sqrt{(D+3)}}      \left\lbrack \theta^{(\mu\nu}\theta^{\rho}_{(\alpha}\theta^{\sigma)}_{\beta}\omega^{\gamma\lambda)} -3\sqrt{2(D+1)}( P^{(0)}_{{21}})^{(\mu\nu\rho\sigma)}_{(\alpha\beta\gamma\lambda)}\right\rbrack. 
	\end{eqnarray}
	It should be noted that the transition operators do satisfy the algebra given by (\ref{algebra}); however, they are not required to complete the identity in (\ref{id}). Nonetheless, they are essential for expanding the sandwiched operator between two rank-4 totally symmetric fields in a bilinear Lagrangian.

	\section{Rank-$(s-4)$ Weyl invariance } \label{app:B}
	
	Since (\ref{weylt}) holds true for any symmetric field, it holds also for the Weyl variation:
	
	\be \delta_{\lambda}\phi (\vf) \equiv \delta_{\lambda}\vf + \sum_{n=2}^{\left\lbrack \frac s2 \right\rbrack }(n-1)c_{n-1} \, \eta^n \, \delta_{\lambda}\vf^{[k]} \quad , \label{deltaphi} \ee
	
	\no in the sense that, by construction, the right side of (\ref{deltaphi}) must be a double traceless tensor. Therefore its $k$-th trace with $k\ge 2$ vanishes identically,

	\be [\delta_{\lambda}\phi (\vf)]^{[k]}=0 \quad ; \quad 2 \le k  \le [s/2] \quad . \label{ktr} \ee
	
	\no On the other hand, we can always rewrite (\ref{deltaphi}) as follows

	\bea [\delta_{\lambda}\phi (\vf)]_{\mu_1\cdots \mu_s} &=&  \eta_{(\mu_1\mu_2}\, \eta_{\mu_3\mu_4}\,[\lambda_{\mu_5\cdots \mu_s)}+ c_1 \, \delta_{\lambda}\phi''_{\mu_5\cdots s)}] + c_2 \,
	 \eta_{(\mu_1\mu_2}\, \eta_{\mu_3\mu_4}\, \eta_{\mu_5\mu_6} \delta_{\lambda}\phi'''_{\mu_7\cdots \mu_s)} + \cdots \nn\\
	 &=& \eta_{(\mu_1\mu_2}\, \eta_{\mu_3\mu_4}\,\left\lbrack \lambda_{\mu_5\cdots \mu_s)}+ c_1 \, \delta_{\lambda}\phi''_{\mu_5\cdots s)} + c_2 \,
	 \eta_{(\mu_5\mu_6}\, \delta_{\lambda}\phi'''_{\mu_7\cdots \mu_s))} + \cdots 
	 \right\rbrack \nn\\ &\equiv & \eta_{(\mu_1\mu_2}\, \eta_{\mu_3\mu_4}\,\tilde{\lambda}_{\mu_5\cdots \mu_s)} \, , \label{lambdatil} \eea
	 
	 \no where $\tilde{\lambda}_{\mu_5\cdots \mu_s}$ is a symmetric rank-$(s-4)$ tensor. By taking traces of (\ref{lambdatil}) backwardly from $k=[s/2]$ until $k=2$ we can show, due to (\ref{ktr}), that $\tilde{\lambda}_{\mu_5\cdots \mu_s} =0$, consequently: $\delta_{\lambda}\phi (\vf) = 0$.

\end{document}